\documentclass{article}
\usepackage{spconf,amsmath,graphicx}
\usepackage{amsbsy}
\usepackage{amssymb}
\usepackage{euscript}
\usepackage{psfrag}
\usepackage{subfigure}
\usepackage{epsfig}

\newtheorem{mytheorem}{\bf Theorem}

\newcommand {\Define} {\stackrel {\Delta} {=}  }

\title{Constant Envelope Precoding for Power-Efficient Downlink Wireless Communication
in Multi-User MIMO Systems Using Large Antenna Arrays}
\name{Saif Khan Mohammed and Erik G. Larsson\thanks{This work was supported by the
    Swedish Foundation for Strategic Research (SSF) and ELLIIT.
    E. G. Larsson is a Royal Swedish Academy of Sciences (KVA)
    Research Fellow supported by a grant from the Knut and Alice
    Wallenberg Foundation.}}
\address{Communication Systems Division, Electrical Eng. (ISY), Link{\"o}ping University, Sweden}
\begin{document}
%\ninept
%
\maketitle
\ninept
\begin{abstract}
We consider downlink cellular multi-user communication between
a base station (BS) having $N$ antennas and $M$ single-antenna users, i.e.,
an $N \times M$ Gaussian Broadcast Channel (GBC).
Under an average only total transmit power constraint (APC), large antenna arrays at the BS (having tens to a few hundred antennas) have been recently shown to achieve remarkable
multi-user interference (MUI) suppression with simple precoding techniques.
However, building large arrays in practice, would require cheap/power-efficient Radio-Frequency(RF)
electronic components. The type of transmitted signal that facilitates the use of
most power-efficient RF components is a {\em constant envelope} (CE) signal (i.e.,
the amplitude of the signal transmitted from each antenna is {\em constant} for every channel use
and every channel realization).
Under certain mild channel conditions (including i.i.d. fading), we analytically show that,
even under the stringent per-antenna CE transmission constraint (compared to APC),
MUI suppression can still be achieved with large antenna arrays.
% i.e., for a fixed $M$, an arbitrarily
%low MUI energy level can be guaranteed by having a sufficiently large $N$.
Our analysis also reveals that,
with a fixed $M$ and increasing $N$, the total transmitted power can be {\em reduced}
while maintaining a constant signal-to-interference-noise-ratio (SINR) level at each user.
We also propose a novel low-complexity CE precoding scheme, using which,
we confirm our analytical observations for the i.i.d. Rayleigh fading channel, through Monte-Carlo simulations.
Simulation of the information sum-rate under the per-antenna CE constraint, shows that, for a fixed $M$
and a fixed desired sum-rate, the required total transmit power
decreases linearly with increasing $N$, i.e.,
an $O(N)$ array power gain.
Also, in terms of the total transmit power required to achieve a fixed desired information sum-rate,
despite the stringent per-antenna CE constraint, the proposed CE precoding scheme performs {\em close} to the GBC sum-capacity (under APC) achieving scheme.
\end{abstract}
\begin{keywords}
GBC, constant envelope, per-antenna.
\end{keywords}
{\vspace{-3mm}}
\section{Introduction}
\label{sec:intro}
We consider a Gaussian Broadcast Channel (GBC), wherein a base station (BS) having $N$ antennas communicates
with $M$ single-antenna users in the downlink.
Large antenna arrays at the BS has been of recent interest, due to their
remarkable ability to suppress multi-user interference (MUI) with very simple precoding techniques.
Specifically, under an average only total transmit power constraint (APC), for a fixed $M$, a simple matched-filter precoder has been shown to achieve total MUI suppression
in the limit as $N \rightarrow \infty$ \cite{TM10}.
Additionally, due to its inherent array power gain property\footnote{\footnotesize{
Under an APC constraint, for a fixed $M$ and a fixed desired information sum-rate, the required total transmit
power decreases with increasing $N$ \cite{DTse}.}},
large antenna arrays are also being considered as an enabler for reducing power consumption
in wireless communications, specially since the operational power consumption at BS is becoming a matter of world-wide
concern \cite{GreenTouch,ComMag2}.
%\footnote{\footnotesize{Operational power consumption at BS has been reported to account for roughly ten percent of the greenhouse
%gas emissions due to ICT \cite{ComMag}}}.

Despite the benefits of large antenna arrays at BS, practically building them would
require cheap and power-efficient RF components like the power amplifier (PA).\footnote{\footnotesize{In conventional BS,
power-inefficient PA's contribute to roughly $40$-$50$ percent of the total operational power consumption \cite{ComMag2}.}}
With current technology, power-efficient RF components are generally non-linear.
The type of transmitted signal that facilitates the use of
most power-efficient/non-linear RF components, is a {\em constant envelope} (CE) signal.
In this paper, we therefore consider a GBC, where the signal transmitted from each BS antenna
has a {\em constant} amplitude for every channel-use and every channel realization.
Since, the per-antenna CE constraint is much more restrictive than APC,
we investigate as to whether MUI suppression and array power gain can still be achieved
under the stringent per-antenna CE constraint ?

To the best of our knowledge, there is no reported work which addresses this question.
Most reported work on per-antenna communication consider an average-only or a peak-only
power constraint (see \cite{WeiYu,KK} and references therein).
In this paper, firstly, we derive expressions for the MUI at each user under the per-antenna CE constraint, and then propose
a low-complexity CE precoding scheme with the objective of minimizing the MUI energy at each user.
For a given vector of information symbols to be communicated to the users, the proposed
precoding scheme chooses per-antenna CE transmit signals in such a way that the MUI energy at each user
is small.

Secondly, under certain mild channel conditions (including i.i.d. fading), using a novel
probabilistic approach, we analytically show that,
{\em MUI suppression can be achieved even under the stringent per-antenna CE constraint.}
Specifically, for a fixed $M$ and fixed user information symbol alphabets, an arbitrarily low
MUI energy can be guaranteed at each user, by choosing a sufficiently large $N$.
Our analysis further reveals that, for i.i.d channels, with a fixed $M$ and increasing $N$, the total transmitted power can be
{\em reduced}
while maintaining a constant SINR level at each user.

Thirdly, through simulation, we confirm our analytical observations for the i.i.d. Rayleigh fading channel.
We numerically compute an achievable ergodic information sum-rate under the per-antenna CE constraint,
and show that, for a fixed $M$ and a fixed desired ergodic sum-rate, {\em the required total transmit power reduces
linearly with increasing $N$}.
We also observe that, to achieve a given desired ergodic information sum-rate, compared to the
optimal GBC sum-capacity achieving scheme under APC, the extra
total transmit power required by the proposed CE precoding scheme is {\em small} (less than $1.7$ dB for large $N$).
\section{System Model}
\label{sec:sysmodel}
Let the complex channel gain between the $i$-th BS antenna and the $k$-th user be denoted
by $h_{k,i}$. The vector of channel gains from the BS antennas to the $k$-th user is
denoted by ${\bf h}_k = (h_{k,1},h_{k,2},\cdots,h_{k,N})^T$.
${\bf H} \in {\mathbb C}^{M \times N}$ is the channel gain matrix with $h_{k,i}$ as its $(k,i)$-th entry.
Let $x_i$ denote the complex symbol transmitted from the $i$-th BS antenna.
Further, let $P_T$ denote the average total power transmitted from all the BS antennas.
Under the APC constraint, we must have ${\mathbb E}[\sum_{i=1}^N \vert x_i \vert^2] = P_T$,
whereas under the per-antenna CE constraint we have $\vert x_i \vert^2 = P_T/N$ which is
clearly a more stringent constraint compared to APC.
Further, due to the per-antenna CE constraint, it is clear that $x_i$ is of the form
$x_i = \sqrt{P_T/N} e^{j \theta_i}$, where $\theta_i$ is the phase of $x_i$.
Under CE transmission, the symbol received
by the users is therefore given by
\begin{eqnarray}
\label{recv_y_k}
y_k = \sqrt{\frac{P_T}{N}} \sum_{i=1}^N h_{k,i} e^{j \theta_i} + w_k \,\,,\,\,k=1,2,\ldots,M
\end{eqnarray}
where $w_k \sim {\mathcal C}{\mathcal N}(0,\sigma^2)$ is the AWGN noise at the $k$-th receiver.
For the sake of notation, let $\Theta = (\theta_1,\cdots,\theta_N)^T$ denote the vector of transmitted phase angles.
Let ${\bf u}= (\sqrt{E_1}u_1,\cdots,\sqrt{E_M}u_M)^T$ be the vector of scaled information symbols, with $u_k \in {\mathcal U}_k$ denoting the information symbol
to be communicated to the $k$-th user. Here ${\mathcal U}_k$ denotes the unit average energy information alphabet of the $k$-th user.
$E_k,k=1,2,\ldots,M$ denote the information symbol energy for each user.
Also, let ${\mathcal U} \Define \sqrt{E_1}{\mathcal U}_1 \times \sqrt{E_2}{\mathcal U}_2 \times \cdots \times \sqrt{E_M}{\mathcal U}_M$.
Subsequently, in this paper, we would be interested in scenarios where $M$ is fixed and $N$
is allowed to increase.
Also, throughout this paper, for a fixed $M$, the alphabets ${\mathcal U}_1,\cdots,{\mathcal U}_M$ are also fixed
and donot change with increasing $N$.
\section{Proposed CE Precoding Scheme}
\label{ce_prec_sec}
%If we adaptively choose ${\mathcal U} \subset {\mathcal M}({\bf H})$ then it is clear that
%for any information symbol vector ${\bf u} \in {\mathcal U}$
%there exists a vector of transmit phase angles $\Theta^u = (\theta_1^u,\theta_2^u,\cdots,\theta_N^u)^T$, such that
%\begin{eqnarray}
%\sqrt{E_k} u_k = \frac{  \sum_{i=1}^N h_{k,i} e^{j \theta_i^u}} {\sqrt{N}}\,\,,\,\, k=1,2,\ldots,M
%\end{eqnarray}
%and therefore, the received noise-free signal at the $k$-th user ($k=1,2,\ldots,M$) i.e.,
%is $\sqrt{P_T} \sqrt{E_k} u_k$, i.e., there is no multi-user interference from information symbols intended
%for the other users.
%
%However, due to difficulty in characterizing the region ${\mathcal M}({\bf H})$ in closed-form,
%subsequently in this paper we assume that, for a given $M$, the alphabet sets ${\mathcal U}_k \,,\,k=1,2,\ldots,M$ are fixed,
%and donot change from one channel realization to another. The alphabet sets also donot change with increasing $N$.
%Also, for a given $M$ and $N$, $E_k$ does not change from one channel realization to another.
%However, for a given $M$ and $N$, based upon performance
%measures like the ergodic information sum rate, the optimal $E_k\,,\,k=1,2,\cdots,M$ is computed a-priori (we shall see this later).
%
For any given information symbol vector ${\bf u}$ to be communicated,
with $\Theta$ as the transmitted phase angle vector, using (\ref{recv_y_k}) the received signal at the $k$-th user can be expressed as
\begin{eqnarray}
\label{yk_model}
y_k & = & \sqrt{P_T} \sqrt{E_k} u_k + \sqrt{P_T} s_k + w_k \nonumber \\
s_k & \Define &  {\Big (} \frac{ \sum_{i=1}^N h_{k,i} e^{j \theta_i}  } { \sqrt{N}} - \sqrt{E_k} u_k   {\Big )}
\end{eqnarray}
where $\sqrt{P_T} s_k$ is the MUI term at the $k$-th user.
For reliable communication to each user, the precoder at the BS, must therefore choose a $\Theta$ such that $\vert s_k \vert$ is as small as possible
for each $k=1,2,\ldots,M$.
This motivates us to consider the following non-linear least squares (NLS) problem
\begin{eqnarray}
\label{NLS}
\Theta^{u} & = & (\theta_1^{u},\cdots,\theta_N^{u})  =  \arg \min_{\theta_i \in [-\pi,\pi), i=1,\ldots,N} g(\Theta,{\bf u}) \nonumber \\
g(\Theta,{\bf u}) & \Define & \sum_{k=1}^M {\Big \vert} \frac{ \sum_{i=1}^N h_{k,i} e^{j \theta_i}  } { \sqrt{N}} - \sqrt{E_k} u_k {\Big \vert}^2.
\end{eqnarray}
This NLS problem is non-convex and has multiple local minima.
However, as the ratio $N/M$ becomes large, due to the large number of extra degrees of freedom ($N - M$),
the value of the objective function $g(\Theta,{\bf u})$ at most local minima has been observed to be small,
enabling gradient descent based methods to be used.
However, due to the slow convergence of gradient descent based methods, we propose a novel iterative method,
which has been experimentally
observed to achieve similar performance as the gradient descent based methods, but with a significantly faster convergence.

In the proposed iterative method to solve (\ref{NLS}), we start with the $p=0$-th iteration,
where we initialize all the angles to $0$.
Each iteration consists of $M$ sub-iterations. Let $\Theta^{(p,q)} = (\theta_1^{(p,q)},\cdots,\theta_N^{(p,q)})^T$
denote the phase angle vector
after the $q$-th sub-iteration ($q=1,2,\ldots,M$) of the $p$-th iteration (subsequently we shall refer to the
$q$-th sub-iteration of the $p$-th iteration as the $(p,q)$-th iteration).
After the $(p,q)$-th iteration, the algorithm moves either to the $(p,q+1)$-th iteration (if $q < M $), or else
it moves to the $(p+1,1)$-th iteration.
In general, in the $(p,q+1)$-th iteration, the algorithm attempts to reduce the current value of the objective function
i.e., ${g}(\Theta^{(p,q)},{{\bf u}})$ by only modifying the $(q+1)$-th phase angle while keeping the other phase angles
fixed to the values from the previous iteration. Therefore, the new phase angles after the $(p,q+1)$-th iteration,
are given by
\begin{eqnarray*}
\hspace{2mm} \theta_{q+1}^{(p,q+1)}  & = & \mbox{arg min}_{_{_{_{\hspace{-12mm}\Theta={\big (}\theta_1^{(p,q)},\cdots,\theta_q^{(p,q)},\phi,\theta_{q+2}^{(p,q)},\cdots,\theta_N^{(p,q)}{\big )}^T \, ,\, \phi \in [-\pi, \pi)}}}} \hspace{-40mm}  {g}(\Theta,{ {\bf u}}) \nonumber \\
= \pi & + & \hspace{-2mm} \arg{\Bigg (} \sum_{k=1}^M \frac {h_{k,q+1}^*}{\sqrt{N}}  {\Big [} \frac{\sum_{_{_{{i=1,\ne (q+1)}}}}^N \hspace{-10mm}  h_{k,i} e^{j \theta_i^{(p,q)}}} {\sqrt{N} } - \sqrt{E_k} {u_k}   {\Big ]}   {\Bigg )} \nonumber \\
\hspace{2mm} \theta_{i}^{(p,q+1)}  & =  & \theta_{i}^{(p,q)} \,\,,\,\,i=1,2,\ldots,N\,,\, i \ne q+1.
\end{eqnarray*}
%The $q+1$-th phase angle is therefore updated as
%{
%\small
%\begin{eqnarray*}
%\theta_{q+1}^{(p,q+1)} = \pi + \arg{\Bigg (} \sum_{k=1}^M \frac {h_{k,q+1}^*}{\sqrt{N}}  {\Big [} \frac{\sum_{_{_{{i=1,\ne (q+1)}}}}^N \hspace{-10mm}  h_{k,i} e^{j \theta_i^{(p,q)}}} {\sqrt{N} } - \sqrt{E_k} {u_k}   {\Big ]}   {\Bigg )}.
%\end{eqnarray*}
%}
The algorithm is terminated after a pre-defined number of iterations.\footnote{\footnotesize{
Experimentally, we have observed that, for the i.i.d. Rayleigh fading channel, with a sufficiently large $N/M$ ratio,
beyond the $p=L N$-th iteration (where $L$ is some constant integer),
the incremental reduction in the value of the objective function is minimal.
Therefore, we terminate at the $L N$-th iteration. Since there are totally $LMN$ sub-iterations,
from the phase angle update equation above, it follows that the complexity of this algorithm is $O(M^2N)$.}}
We denote the phase angle vector after the last iteration by
${\widehat \Theta^{u}} = ({\widehat \theta_1^{u}},\cdots,{\widehat \theta_N^{u}} )^T$.

With ${\widehat \Theta^{u}}$ as the transmitted phase angle vector,
the received signal-to-noise-and-interference-ratio (SINR) at the $k$-th user is given by
%\begin{eqnarray}
%\label{s_k_hat}
%y_k
%& = & \sqrt{P_T} \sqrt{E_k} u_k + \sqrt{P_T} {\widehat s_k} + w_k \,,\, k=1,2,\ldots,M \nonumber \\
%{\widehat s_k} & \Define &  {\Big (} \frac{ \sum_{i=1}^N h_{k,i} e^{j {\widehat \theta_i^u} }  } { \sqrt{N}} - \sqrt{E_k} u_k   {\Big )}
%\end{eqnarray}
%The average signal-to-noise-and-interference-ratio (SINR) at the users is then given by
{
\vspace{-2mm}
}
\begin{eqnarray}
\label{sinr_eq}
\gamma_k({\bf H},E) & = &\frac { E_k} { {\mathbb E}_{_{u_1,\cdots,u_M}}{\big [} {\vert}  {\widehat s_k} { \vert}^2 {\big ]}  + \frac{\sigma^2}{P_T} } \nonumber \\
{\widehat s_k} & \Define &  {\Big (} \frac{ \sum_{i=1}^N h_{k,i} e^{j {\widehat \theta_i^{u}} }  } { \sqrt{N}} - \sqrt{E_k} u_k   {\Big )}
\end{eqnarray}
where $E \Define (E_1,E_2,\cdots,E_M)^T$ is the vector of information symbol energy.
%From (\ref{sinr_eq}), it is clear that, for a fixed $P_T/\sigma^2$, the average SINR at the $k$-th user is dependent upon the mean energy of the
%multi-user interference term ${\widehat s_k}$, as well as $E_k$.
For each user, we would be ideally interested to have a low value of the MUI energy ${\mathbb E}[ \vert {\widehat s_k} \vert^2]$, since this would
imply a larger SINR.
%Based on (\ref{sinr_eq}), we have the following two important remarks which are formally shown to be true in Sections \ref{dim_sec} and \ref{inc_sec}.
{
\vspace{-3mm}
}
\section{MUI Analysis}
\label{muianalysis}
In this Section, for any general CE precoding scheme (without restricting to the proposed CE precoding algorithm in Section \ref{ce_prec_sec}),
through analysis, we aim to get a better understanding of the MUI energy level at each user. Towards this end, we firstly
study the dynamic range of values taken by the noise-free received signal
at the users, which is given by the set
\begin{eqnarray}
\label{MH_def}
{\mathcal M}({\bf H})  & \Define &  {\Big \{} {\bf v}=(v_1,\cdots,v_M) \,\,{\big |}\,\,  \nonumber \\
& & { v_k} = \frac {\sum_{i=1}^N h_{k,i} e^{j \theta_i}}{\sqrt{N}} \,,\,\theta_i \in [-\pi,\pi) {\Big \}} 
%{\bf e}(\Theta) & \Define & (e^{j \theta_1}, e^{j \theta_2},\cdots, e^{j \theta_N})^T.
\end{eqnarray}
For any vector ${\bf v} \in {\mathcal M}({\bf H})$, from (\ref{MH_def})
it follows that there exists a $\Theta^v$ such that ${ v_k} = \frac {\sum_{i=1}^N h_{k,i} e^{j \theta_i^v}}{\sqrt{N}}$.
This sum can now be expressed as a sum of
$N/M$ terms (without loss of generality let us assume that $N/M$ is integral only for the argument
presented here)
\begin{eqnarray}
\label{eq_sum}
v_k  =  \sum_{q=1}^{N/M} v_k^q \,\,\,,\,\,\,
v_k^q \Define \hspace{-6mm}  \sum_{r=(q-1)M+1}^{qM} \hspace{-5mm} h_{k,r} e^{j \theta_r^v} \,,\,q=1,\ldots,\frac{N}{M}.
\end{eqnarray}
From (\ref{eq_sum}) it immediately follows that ${\mathcal M}({\bf H})$ can be expressed
as a direct-sum of $N/M$ sets, i.e.
\begin{eqnarray}
{\mathcal M}({\bf H}) & = & {\mathcal M}({\bf H}^{1}) \oplus {\mathcal M}({\bf H}^{2}) \oplus \cdots \oplus {\mathcal M}({\bf H}^{N/M}) \nonumber \\
{\mathcal M}({\bf H}^{q}) & \Define & {\Big \{} {\bf v}=(v_1,\cdots,v_M) \,\,{\big |}\,\, \nonumber \\
& & \hspace{-10mm}  { v_k} = \frac {\sum_{i=1}^M h_{k,(q-1)M+i} \hspace{3mm} e^{j \theta_i}}{\sqrt{N}} \,,\,\theta_i \in [-\pi,\pi) {\Big \}} 
\end{eqnarray}
where ${\mathcal M}({\bf H}^{q}) \subset {\mathbb C}^M$ is the dynamic range of the received noise-free signals
when only the $M$ BS antennas numbered $(q-1)M+1,(q-1)M+2,\cdots,qM$ are used and the remaining $N - M$ antennas are inactive.
If the statistical distribution of the channel gain vector from a BS antenna to all the users is identical for all the BS antennas,
then, on an average the sets ${\mathcal M}({\bf H}^{q})$ would have have similar topological properties.
Since, ${\mathcal M}({\bf H})$ is a direct-sum of $N/M$ topologically similar sets, it is expected that for a fixed $M$, on an average the
region ${\mathcal M}({\bf H})$ expands/enlarges with increasing $N$.
Based on this discussion, for i.i.d. channels, we have the following two important remarks in Section \ref{remark1_sec} and \ref{remark2_sec}.
{\vspace{-2mm}}
\subsection{Diminishing MUI with increasing $N$, for fixed $M$ and $E_k$}
\label{remark1_sec}
\begin{figure}[t]
\begin{center}
\epsfig{file=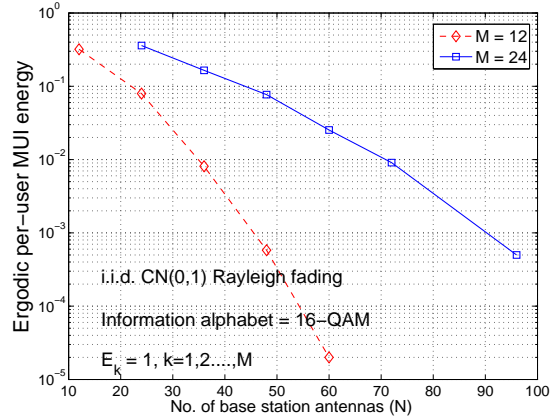, width=75mm,height=59mm}
\end{center}
\vspace{-6mm}
\caption{Reduction in MUI with increasing $N$. Fixed $E_k$.}
\label{fig_0}
\vspace{-4mm}
\end{figure}

\begin{mytheorem}\label{main_theorem1}
For a fixed $M$ and increasing $N$, consider a sequence of channel gain matrices $\{ {\bf H}_N \}_{N=M+1}^{\infty} $ satisfying the mild conditions
{
\vspace{-4mm}
}
\begin{eqnarray}
\label{assumptions}
\hspace{-19mm} \lim_{N \rightarrow \infty} \frac{\vert {{\bf h}_{k}^{(N)}}^H {\bf h}_{l}^{(N)}\vert } {N }  =  0 \,\,,\,\,\forall \, k \ne l \,\, \mbox{(cnd.1)} \nonumber \\
\lim_{N \rightarrow \infty} \frac {\sum_{i=1}^N  \vert h_{{k_1},i} \vert \vert h_{{l_1},i} \vert  \vert h_{{k_2},i} \vert \vert h_{{l_2},i} \vert }
 {  N^2 }   =  0 \nonumber \\
\,\,,\,\,\forall k_1,l_1,k_2,l_2 \in (1,2,\ldots,M) \,\, & & \mbox{(cnd.2)} \nonumber \\
\lim_{N \rightarrow \infty} \frac{\Vert {\bf h}_{k}^{(N)} \Vert^2  } {N }  =  c_k \,\,,\,\, k=1,2,\ldots,M \,\,  \mbox{(cnd.3)}
\end{eqnarray}
where $c_k$ are positive constants and ${\bf h}_{k}^{(N)}$ denotes the $k$-th row of ${\bf H}_N$.
(From the law of large numbers, it follows that i.i.d. channels satisfy these conditions with probability $1$.)

For any given fixed finite alphabet ${\mathcal U}$ (fixed $E_k,k=1,\ldots,M$) and any given $\Delta > 0$, there exists a corresponding integer $N(\{ {\bf H}_{N} \}, {\mathcal U}, \Delta)$
such that with $N \geq N(\{ {\bf H}_{N} \}, {\mathcal U}, \Delta)$ and ${\bf H}_N$ as the channel gain matrix, for any ${\bf u} \in {\mathcal U}$ to be communicated,
there exist a phase angle vector
$\Theta_N^u(\Delta)$ = $(\theta_1^u(\Delta),\cdots,\theta_N^u(\Delta))^T$ which when transmitted,
results in the MUI energy at each user being upper bounded by $2 \Delta^2$, i.e.
{\vspace{-2mm} }
\begin{eqnarray}
\label{thm1_eqn}
{\Big \vert } \frac{ \sum_{i=1}^N h_{k,i}^{(N)} e^{j \theta_i^u(\Delta)}}   { \sqrt{N}} - \sqrt{E_k} u_k   {\Big \vert }^2 \leq 2 \Delta^2 \,\,,\,\,k=1,\ldots,M
\end{eqnarray}
where $h_{k,i}^{(N)}$ denotes the $i$-th component of ${\bf h}_{k}^{(N)}$.
\end{mytheorem}
Due to limited space, we present a sketch of the proof of Theorem \ref{main_theorem1} in Appendix \ref{dim_sec}.
In Theorem \ref{main_theorem1}, 
$\Delta$ can be chosen to be arbitrarily small,
and therefore,
the {\em MUI energy at each user can be guaranteed to be arbitrarily small, by choosing a sufficiently large $N$}.
In Fig.~\ref{fig_0}, for the i.i.d. ${\mathcal C}{\mathcal N}(0,1)$ Rayleigh fading channel,
with fixed information alphabets
${\mathcal U}_1 = {\mathcal U}_2 = \cdots = {\mathcal U}_M$ = $16$-QAM and
fixed information symbol energy $E_k=1,k=1,\ldots,M$,
we plot the ergodic (averaged w.r.t. channel statistics) MUI energy ${\mathbb E}_{\bf H}[\vert {\widehat s_k} \vert^2]$ (observed to be same for each user)
as a function of increasing $N$ (${\widehat s_k}$ is given by (\ref{sinr_eq})).
It is observed that, for a fixed $M$, fixed information alphabets and fixed information symbol energy, the ergodic per-user MUI energy decreases with increasing $N$.
%Since the MUI energy is plotted on a log scale, the linear slope of the ergodic MUI energy curves suggest that the ergodic MUI energy decreases
%exponentially in $N$.
\begin{figure}[t]
\begin{center}
\epsfig{file=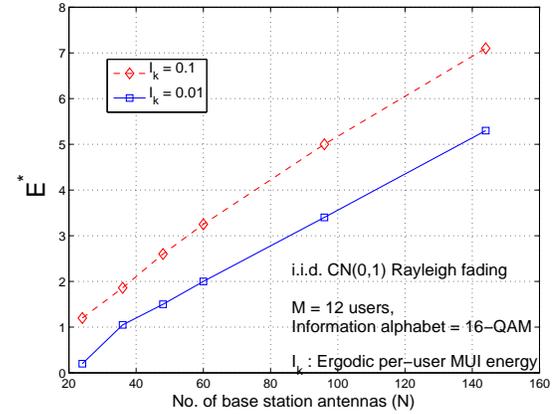, width=75mm,height=59mm}
\end{center}
\vspace{-6mm}
\caption{$E^\star$ vs. $N$. Fixed MUI energy (same for each user).}
\label{fig_1}
\vspace{-5mm}
\end{figure}
{\vspace{-3mm}}
\subsection{Increasing $E_k$ with increasing $N$, for a fixed MUI}
\label{remark2_sec}
%In (\ref{sinr_eq}), with increase in $E_k$, the numerator of the SINR expression increases, but at the same time
%multi-user interference energy term in the denominator may also increase if $E_k$ is so large that
%$(\sqrt{E_1}u_1,\cdots,\sqrt{E_M}u_M)^T$ lies outside the region ${\mathcal M}({\bf H})$.
%There is therefore a tradeoff between the information symbol energy and the energy of the multi-user interference term.
%This then suggests that there exists an optimal value of $E_k\,,\,k=1,2,\cdots,M$ with respect to maximizing the
%average SINR for all the users.
%From Remark 1, we know that with increasing $N$ but fixed $E_k$ ($k=1,2,\cdots,M$), the multi-user interference energy decreases with high probability.
%Therefore, it can be argued that, since the set ${\mathcal M}({\bf H})$ enlarges with increasing $N$, for a fixed desired level of multi-user interference energy, the
%symbol energy $E_k (k=1,2,\cdots,M$) can be increased with increasing $N$.
%
%Since, with increasing $N$ and fixed $E_k\,,\,k=1,2,\cdots,M$, an arbitrarily small multi-user interference energy can be guaranteed at each user,
%it can be argued that, with increasing $N$ the information symbol energy $E_k\,,\,k=1,2,\cdots,M$ can be increased while maintaining a fixed multi-user interference energy level at each user.
From (\ref{sinr_eq}), it is clear that, for a fixed $M$ and $N$, increasing $E_k,k=1,\ldots,M$ would enlarge ${\mathcal U}$ which could
then increase MUI energy level at each user. However, since increase in $N$ results in reduction of MUI (Theorem \ref{main_theorem1}),
it can be argued that, with increasing $N$ the information symbol energy
of each user can be increased while still maintaining a fixed MUI energy level at each user.
%In Section \ref{inc_sec}, through simulations, we show that indeed, for the i.i.d. Rayleigh fading channel with fixed $M$, $E_k$ can be increased linearly with increasing $N$,
%while still maintaining a fixed MUI energy level at each user.
We illustrate this through the following example.
Let the fixed desired ergodic MUI energy level for the $k$-th user be denoted by
$I_k\,,\,k=1,2,\cdots,M$. For the sake of simplicity we consider ${\mathcal U}_1 = {\mathcal U}_2 = \cdots = {\mathcal U}_M$.
Consider the following optimization
{\vspace{-3mm}}
\begin{eqnarray}
\label{e_star}
E^\star \Define  \arg \hspace{-5mm} \max_{_{p > 0 \,\,{\big |}\,\, E_k = p \,,\, I_k = {\mathbb E}_{{\bf H}}{\big [}{\mathbb E}_{_{u_1,\cdots,u_M}}{\big [}
{\vert} {\widehat s_k}  {\vert}^2 {\big ]}{\big ]} \,\,,\,\,k=1,\cdots,M}}  \hspace{-20mm}  p
\vspace{-3mm}
\end{eqnarray}
which finds the highest possible equal energy of the information symbols under the constraint that the ergodic MUI
energy level is fixed at $I_k\,,\,k=1,2,\cdots,M$.
%The fact that we choose same information symbol energy for each user is not restrictive since all users have the same
%information alphabet and channel fading is i.i.d. Rayleigh.
In (\ref{e_star}),
${\widehat s_k}$ is given by (\ref{sinr_eq}).
In Fig.~\ref{fig_1}, for the i.i.d. Rayleigh fading channel,
for a fixed $M = 12$ and a fixed ${\mathcal U}_1=\cdots={\mathcal U}_M=16$-QAM,
we plot $E^\star$ as a function of increasing $N$, for two different fixed desired MUI energy
levels, $I_k= 0.1$ and $I_k=0.01$ (same $I_k$ for each user).
(Due to same channel distribution and information alphabet for each user,
it is observed that the ergodic MUI energy level is also same if the users have equal information symbol energy.)
From Fig.~\ref{fig_1}, it can be observed that for a fixed $M$ and fixed ${\mathcal U}_1,\cdots,{\mathcal U}_M$, indeed,
$E^\star$ increases linearly with increasing $N$, while still maintaining a fixed MUI energy level at each user.
At low MUI energy levels, from (\ref{sinr_eq}) it follows that $\gamma_k \approx P_T E_k /\sigma^2$.
Since $E_k$ ($k=1,2,\cdots,M$) can be increased linearly with $N$ (while still maintaining low MUI level),
it can be argued that a desired fixed SINR level can be maintained at each user by simply {\em reducing}
$P_T$ linearly with increasing $N$.
%This then suggests that, for a fixed $M$, an array power gain of $O(N)$ can be achieved under the stringent per-antenna CE constraint.
%In Section \ref{apg}, we confirm this observation on the achievability of an $O(N)$ array power gain.
This suggests the achievability of an $O(N)$ array power gain.
{\vspace{-3mm}}
\section{Achievable array power gain }
\label{apg}
\begin{figure}[t]
\begin{center}
\epsfig{file=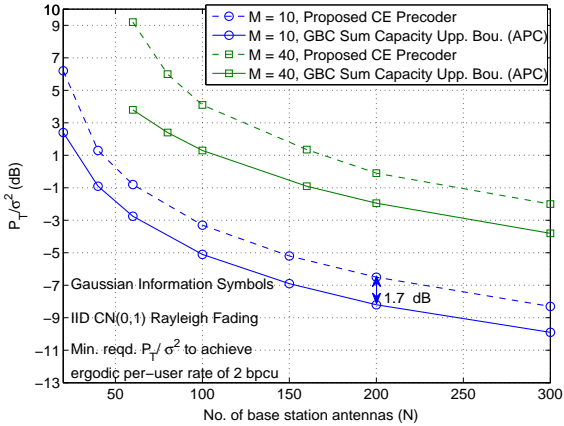, width=75mm,height=60mm}
\end{center}
{\vspace{-6mm}}
\caption{Reqd. $P_T/\sigma^2$ vs. $N$. Fixed ergodic per-user rate $=2$ bpcu.}
{\vspace{-4mm}}
\label{fig_2}
{\vspace{-2mm}}
\end{figure}

{\vspace{-2mm}}
%For a given ${\bf H}$,
%,$E = (E_1,E_2,\cdots,E_M)^T$,
%and Gaussian information alphabets, it can be shown that 
%under the per-antenna CE constraint an information rate of
%$\log_2(\gamma_k({\bf H},E))$ is achievable for the $k$-the user ($\gamma_k({\bf H},E)$ is given by (\ref{sinr_eq})).
%\begin{eqnarray}
%\label{mi_expr}
%R_k({\bf H},E) & \Define & I(r_k ; {u_k}) \geq \log_2(\gamma_k({\bf H},E))
%\end{eqnarray}
%for each $k=1,2,\cdots,M$, where $\gamma_k({\bf H},E)$ is given by (\ref{sinr_eq}).
%With $ \log_2(\gamma_k({\bf H},E))$ as an achievable ergodic rate for the $k$-th user,
For the proposed CE precoding scheme in Section \ref{ce_prec_sec}, with the same Gaussian information alphabet for each user, an achievable ergodic information sum-rate for GBC under the per-antenna CE constraint, can be shown to be given by
$R_{\mbox{\footnotesize{CE}}}(E) = \sum_{k=1}^M {\mathbb E}_{{\bf H}} \log_2(\gamma_k({\bf H},E))  $ (Here we
have used the fact that, with Gaussian alphabet, Gaussian noise is the worst noise in terms of
achievable mutual information).
%The optimal vector of information symbol energy is then defined as
%\begin{eqnarray}
%\label{E_opt}
%E^{\mbox{\footnotesize{opt}}} \Define \arg \max_{E \in {\mathbb R^+}^M} R_{\mbox{\footnotesize{sum}}}(E).
%\end{eqnarray}
%Based on Remark 2, we would be interested in maximizing $R_{\mbox{\footnotesize{sum}}}(E)$ w.r.t. $E$ (an $M$-dimensional optimization).
%is numerically complex for large $M$).
%For the i.i.d. fading channel,
%since the channel gains are i.i.d. and the information alphabet is same for each user, it is highly
%likely that the optimal symbol energy level for each user is the same due to symmetry.
We numerically optimize $R_{\mbox{\footnotesize{CE}}}(E)$ subject to the constraint
$E_1=\cdots=E_M$.
Based on this optimized ergodic sum-rate, in Fig.~\ref{fig_2}, for the i.i.d. ${\mathcal C}{\mathcal N}(0,1)$ Rayleigh fading channel,
we plot the required $P_T/\sigma^2$ to achieve an ergodic per-user
information rate of $2$ bits-per-channel-use (bpcu) (We have observed that the ergodic information rate achieved by each user is $1/M$ of the ergodic sum-rate). It is observed that, for a fixed $M$, at sufficiently large $N$, the required $P_T/\sigma^2$
reduces by roughly $3$ dB for every doubling in $N$
(i.e., the required $P_T/\sigma^2$ reduces linearly with increasing $N$). 
{\em This shows that, for a fixed $M$, an array power gain of $O(N)$ can indeed be achieved even under the stringent
per-antenna CE constraint.}
For the sake of comparison, we have also plotted
the minimum $P_T/\sigma^2$ required under the APC constraint (we have used the co-operative upper bound on the GBC sum-capacity \cite{SV}).
We observe that, for large $N$ and a fixed per-user desired ergodic information rate of $2$ bpcu, compared to the APC only constrained GBC,
the extra total transmit power required under the more stringent per-antenna CE constraint is {\em small} (only $1.7$ dB).
\bibliographystyle{IEEEbib}
\bibliography{strings,refs}
\appendix
{\vspace{-3mm}}
\section{Proof of Theorem {\bf 1} (Sketch)}
\label{dim_sec}
Let us consider a probability space 
with the transmitted phase angles $\theta_i,i=1,2,\ldots,N$ being i.i.d.
r.v's uniformly distributed in $[-\pi \,,\, \pi)$.
For a given sequence of channel matrices $\{ {\bf H}_N \}$, we define a corresponding sequence of r.v's $\{ {\bf z}_N \}$, with
${\bf z}_N \Define (z_{1}^{I^{(N)}},z_{1}^{Q^{(N)}},\ldots,z_{M}^{I^{(N)}},z_{M}^{Q^{(N)}}) \in {\mathbb R}^{2M}$, where we have
$z_{k}^{I^{(N)}} \Define$ $\mbox{Re}{\Big (} \frac{\sum_{i=1}^M h_{k,i}^{(N)} e^{j \theta_i}} {\sqrt{N}}  {\Big )}$,
$z_{k}^{Q^{(N)}} \Define \mbox{Im}{\Big (} \frac{\sum_{i=1}^M h_{k,i}^{(N)} e^{j \theta_i}} {\sqrt{N}}  {\Big )}$, $k=1,\ldots,M$.
Using the Lyapunov Central Limit Theorem (CLT) \cite{Patrick}, it can be shown that, for any channel sequence $\{ {\bf H}_N \}$ satisfying the
conditions in (\ref{assumptions}), as $N \rightarrow \infty$, the corresponding sequence of r.v's $\{ {\bf z}_N \}$ 
converges
in distribution to a $2M$-dimensional real Gaussian random vector $X = (X_1^I,X_1^Q,\cdots,X_M^I,X_M^Q)^T$ with independent zero-mean
components and $\mbox{var}(X_k^I) = \mbox{var}(X_k^Q) = c_k/2$.

This then implies, that, asymptotically as $N \rightarrow \infty$, the limiting range space of the random variable ${\bf z}_N$
is the whole space ${\mathbb R}^{2M}$.\footnote{\footnotesize{
Further, for any
value taken by ${\bf z}_N$, say ${\bf z}_N={\bf e}=(e_1^I,e_1^Q,\cdots,e_M^I,e_M^Q)^T$, from the definition of ${\mathcal M}({\bf H})$ in (\ref{MH_def}), it is clear that
the complex vector ${\bf e}^c = (e_1^c,e_2^c,\cdots,e_M^c)^T$ ($e_k^c \Define (e_k^I + j e_k^Q)$) belongs to ${\mathcal M}({\bf H}_N)$.
Hence, it can be concluded that ${\mathcal M}({\bf H}_N) \rightarrow {\mathbb C}^M$ as $N \rightarrow \infty$ (supports our observation
on the enlargement of ${\mathcal M}({\bf H})$ with increasing $N$).}}
For a given ${\bf u} \in {\mathcal U}$, and $\Delta > 0$, we next consider the box
\begin{eqnarray}
\label{box_eqn}
{\mathcal B}_{_{\Delta}}({\bf u}) & \Define & {\Big \{} {\bf b}=(b_1^I,b_1^Q,\cdots,b_M^I,b_M^Q)^T \in {\mathbb R}^{2M} \,|\, \nonumber \\
& & \hspace{-19mm} \vert b_k^I - \sqrt{E_k} u_k^I \vert \leq \Delta \,,\, \vert b_k^Q - \sqrt{E_k} u_k^Q \vert \leq \Delta\,\,,\,\,k=1,2,\ldots,M     {\Big \}} \nonumber \\
& & u_k^I \Define \mbox{Re}(u_k) \,,\, u_k^Q \Define \mbox{Im}(u_k) 
\end{eqnarray}
The box ${\mathcal B}_{_{\Delta}}({\bf u})$
contains all those vectors in ${\mathbb R}^{2M}$ whose component-wise displacement from ${\bf u}$ is upper bounded by $\Delta$.
Using the fact that ${\bf z}_N$ converges in distribution to a Gaussian r.v. with ${\mathbb R}^{2M}$ as its range space, continuity arguments show that,
for any $\Delta > 0$, there exist an integer $ N(\{ {\bf H}_N \},{\mathcal U}, \Delta)$, such that
for all $N \geq N(\{ {\bf H}_N \},{\mathcal U}, \Delta)$
\begin{eqnarray}
\mbox{Prob}({\bf z}_N \in {\mathcal B}_{_{\Delta}}({\bf u})) > 0 \,\,,\,\, \forall \, {\bf u} \in {\mathcal U}.
\end{eqnarray}
Since, the probability that ${\bf z}_N$ lies in the box ${\mathcal B}_{_{\Delta}}({\bf u})$, is {\em strictly positive},
it follows that there exist a phase angle vector $\Theta_N^u(\Delta)=(\theta_1^u(\Delta),\cdots,\theta_N^u(\Delta))^T$ which satisfies (\ref{thm1_eqn}).
\end{document}